# A fluorescent-protein spin qubit

Jacob S. Feder*[1], Benjamin S. Soloway*[1], Shreya Verma[2], Zhi Z. Geng[1], Shihao Wang[1], Bethel Kifle[1], Emmeline G. Riendeau[3], Yeghishe Tsaturyan[1], Leah R. Weiss[1], Mouzhe Xie[1,4], Jun Huang[1], Aaron Esser-Kahn[1,5], Laura Gagliardi[1,2,6], David D. Awschalom[†1,3,7], & Peter C. Maurer[†1,5,7]

[1] Pritzker School of Molecular Engineering, University of Chicago, Chicago, IL 60637, USA.
[2] Department of Chemistry, University of Chicago, Chicago, IL 60637, USA.
[3] Department of Physics, University of Chicago, Chicago, IL 60637, USA.
[4] School of Molecular Sciences, Arizona State University, Tempe, AZ 85281, USA.
[5] CZ Biohub Chicago, LLC, Chicago, IL 60642, USA.
[6] James Franck Institute, University of Chicago, Chicago, IL 60637, USA.
[7] Center for Molecular Engineering and Materials Science Division, Argonne National Laboratory, Lemont, IL 60439, USA.
*These authors contributed equally
[†] awsch@uchicago.edu and pmaurer@uchicago.edu

**Optically-addressable spin qubits form the foundation of a new generation of emerging nanoscale sensors [1–5]. The engineering of these sensors has mainly focused on solid-state systems such as the nitrogen-vacancy (NV) center in diamond. However, NVs are restricted in their ability to interface with biomolecules due to their bulky diamond host. Meanwhile, fluorescent proteins have become the gold standard in bioimaging, as they are genetically encodable and easily integrated with biomolecules [6–8]. While fluorescent proteins have been suggested to possess a metastable triplet state [9], they have not been investigated as qubit sensors. Here, we realize an optically-addressable spin qubit in the enhanced yellow fluorescent protein (EYFP) enabled by a novel spin-readout technique. A near-infrared laser pulse allows for triggered readout of the triplet state with up to 44% spin contrast. Using coherent microwave control of the EYFP spin at liquid-nitrogen temperatures, we measure a spin-lattice relaxation time of $(141\pm5)$ μs, a $(16\pm2)$ μs coherence time under Carr-Purcell-Meiboom-Gill (CPMG) decoupling, and predict an oscillating (AC) magnetic field sensitivity with an upper bound of 183 fT $mol^{1/2}$ $Hz^{-1/2}$. We express the qubit in mammalian cells, maintaining contrast and coherent control despite the complex intracellular environment. Finally, we demonstrate optically-detected magnetic resonance at room temperature in aqueous solution with contrast up to 3%, and measure a static (DC) field sensitivity with an upper bound of 93 pT $mol^{1/2}$ $Hz^{-1/2}$. Our results establish fluorescent proteins as a powerful new qubit sensor platform and pave the way for applications in the life sciences that are out of reach for solid-state technologies.**

Optically-addressable quantum sensors capable of measuring nanoscale magnetic fields [10–12], electric fields [13], and temperature [14–16] have had a lasting impact on the physical sciences [1, 17–20]. In contrast, their adoption in the life sciences has been limited, with most applications remaining at the proof-of-concept stage. A molecule-scale spin qubit sensor that can be readily interfaced with biological target systems could enable a new generation of ultra-sensitive measurement techniques for fundamental research and medical diagnostics [3, 4].

The leading spin qubit for biological sensing is the nitrogen-vacancy (NV) center in diamond because it has a spin that can be optically initialized and read out, remains coherent at room temperature, and can be hosted in diamond nanoparticles. Nanodiamonds suffer from problems common to nanoparticle labeling [21] since they are heterogeneous in size and morphology, have complex surface chemistry, and are ten times larger than an average protein [22, 23]. These issues have made cellular delivery and specific targeting an outstanding challenge. Optically-addressable molecular spin qubits, such as polycyclic aromatic hydrocarbons [24–27], metal-organic complexes [28–30], and radical systems [31, 32], offer potential advantages over NV centers. However, the systems examined thus far each have their own set of shortcomings, which include low spin contrast, poor photon-emission rates, lack of water solubility, the requirement of a solid-state host, or reactivity with ambient air, limiting their applicability for most biosensing applications.

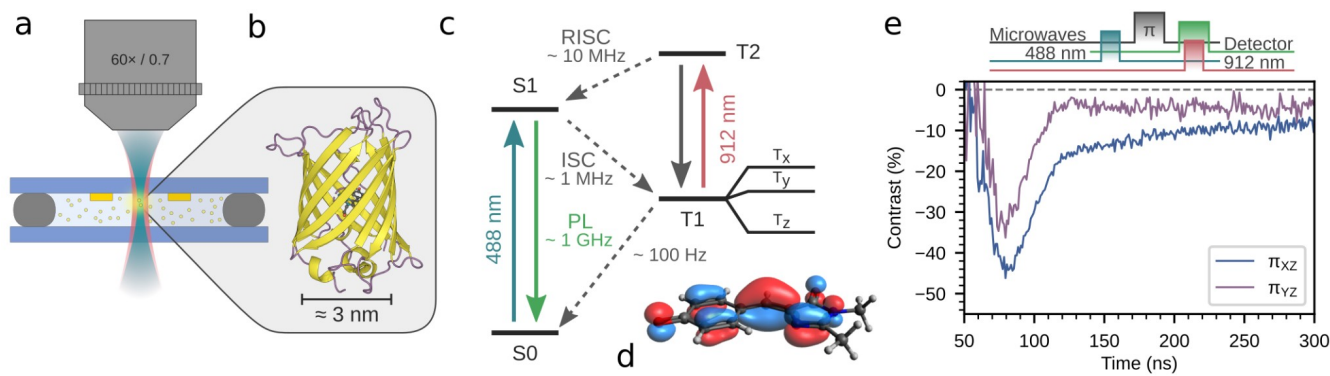

**Fig. 1 | Photophysics of EYFP proteins and OADF readout scheme. a,** Schematic representation of purified optically-addressable EYFP spin qubits (yellow) encapsulated in a liquid cell, which is composed of sapphire coverslips (blue) sealed with an o-ring (dark gray). The sample is imaged in a custom-built confocal microscope (Extended Data Fig. 1). A microscope objective (gray) focuses gated laser beams of 488 nm (light blue) and 912 nm (red) wavelengths to an approximately 5 μm diameter focal spot. A lithographically patterned microwave structure (gold) is used for coherent driving of the EYFP spins. **b,** Ribbon diagram of EYFP (Protein Data Bank: 2YFP [51]). **c,** Energy-level diagram of EYFP with OADF readout scheme. **d,** The natural transition orbital structure for the T1 state, calculated by *ab initio* calculations. **e,** Contrast of OADF-based spin readout as a function of time after the 912 nm laser pulse for the $T_x$-$T_z$ (blue) and $T_y$-$T_z$ (purple) transition. The measurement sequence above the figure illustrates the timing of the microwave and optical pulses. The "Detector" channel represents the timing of the acousto-optic modulator used to gate the detector. Panel **e** is measured at 80 K.

Fluorescent proteins derived from the jellyfish *Aequorea victoria* are a ubiquitous tag for biological imaging [6–8]. They have a high absorption cross section and quantum yield, and a short fluorescence lifetime, resulting in bright emission. The ability to be genetically encoded makes them ideal for dense labeling of living cells and organisms, with specificity beyond what can be achieved with synthetic molecular fluorophores or nanoparticles [8]. Decades of research has generated a vast library of fusion proteins consisting of fluorescent proteins deterministically tagged to biological targets. Interestingly, the photoactive organic fluorophore embedded in the fluorescent proteins' beta barrel possesses a metastable state that has been suggested to be a spin triplet [9]. If harnessed as a qubit, this intrinsic spin could unlock new quantum sensing modalities for fluorescent proteins.

**Protein qubit and spin readout**

Here, we use a custom confocal microscope (Fig. 1a) to optically address the metastable triplet spin state of the EYFP (Fig. 1b). The EYFP spin is initialized with a 488 nm optical pulse, cycling the fluorophore between its ground (S0) and first-excited (S1) singlet state, which then undergoes intersystem crossing (ISC) to a spin-polarized metastable triplet (T1). Time-dependent density functional theory (TDDFT) calculations show that the T1 spin density is fully delocalized on the chromophore (Fig. 1d). However, this metastable triplet presents challenges in its use as a spin qubit. Readout of the spin via the triplet decay is precluded by spin depolarization at most temperatures, which is fast compared to the millisecond-scale triplet lifetime [9, 33]. The extended triplet lifetime also limits the measurement sequence repetition rate, placing a bottleneck on the sensitivity.

We overcome the aforementioned challenges associated with the metastable triplet by utilizing a photophysical technique to shorten its lifetime [34, 35] and enable on-demand spin readout (Fig. 1c). A 912 nm optical pulse drives the fluorophore from its triplet T1 state to a higher-lying triplet. TDDFT calculations (Methods) suggest the identity of this triplet to be the T2 state. Spin-dependent reverse intersystem crossing (RISC) to the singlet manifold is accelerated in this state. Following RISC, a background-free delayed fluorescence photon is emitted from S1. Readout using this optically activated delayed fluorescence (OADF) [34] scheme is approximately five orders of magnitude faster than via the metastable triplet decay.

We find that the OADF spectrum closely resembles the standard fluorescence spectrum at 80 K (Extended Data Fig. 3a), which indicates that the delayed fluorescence photons are indeed emitted from the S1 to S0 transition. Furthermore, the wavelength of the intensity maxima at 80 K and room temperature are in good agreement, suggesting that the frozen protein maintains its structure in the vicinity of the fluorophore after being slowly cooled to cryogenic temperatures (Extended Data Fig. 2a, b).

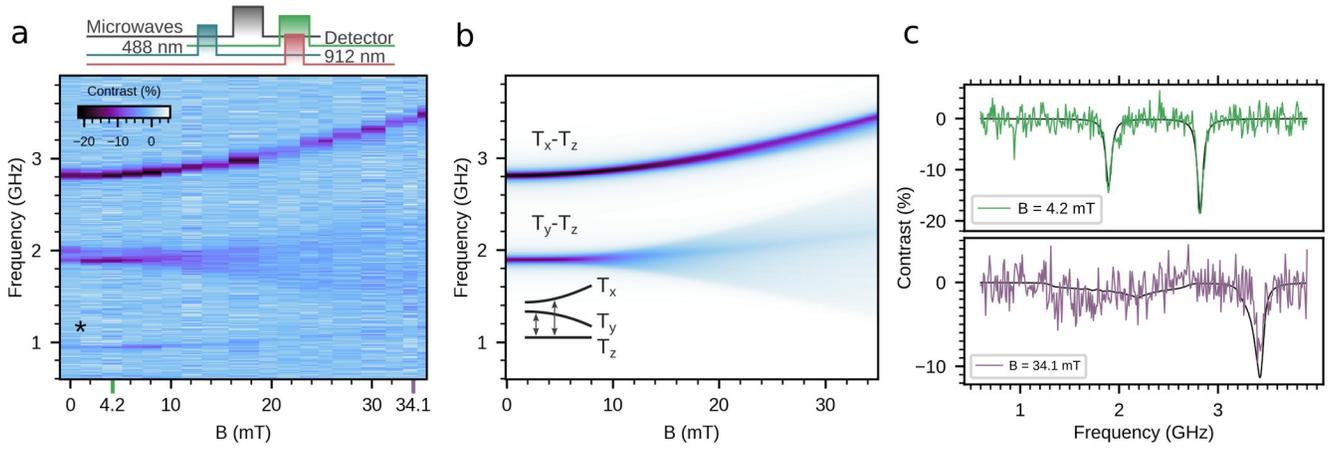

**Fig. 2 | ODMR spectroscopy of EYFP. a,** Experimental ODMR signal as a function of externally applied magnetic field. The faint resonance (*) at approximately $\omega = (2\pi) \times 0.9$ GHz is associated with microwave harmonics that drive the $T_x$-$T_z$ and $T_y$-$T_z$ transitions. **b,** Simulated ODMR response based on the model from equation 1. The fit parameters (D, E, linewidth of resonance, and $T_x$-$T_z$ and $T_y$-$T_z$ amplitudes) are extracted from an ODMR measurement at low magnetic fields (Methods). The ODMR amplitude for the $T_x$-$T_y$ transition is set to 0. For the resonance frequencies of the diagonalized Hamiltonian, see Extended Data Figure 4. **c,** Experimental (green, purple) and simulated (black) ODMR spectrum at 4.2 mT and 34.1 mT. Panels **a**, **c** are measured at 80 K.

The addition of a microwave drive demonstrates the optical addressability of the spin state. After initialization, a microwave $\pi$-pulse induces population transfer between the spin sublevels in the T1 state. Detection of the OADF photons triggered by the 912 nm laser pulse yields a maximum spin contrast of 44% for the $T_x$-$T_z$ transition and 32% for the $T_y$-$T_z$ transition at 80 K (Fig. 1e, Extended Data Fig. 3b,c).

**Optically-detected magnetic resonance**

We investigate the nature of the metastable triplet spin state through optically-detected magnetic resonance (ODMR) spectroscopy. Figure 2a shows the ODMR contrast as a function of microwave frequency and external magnetic field. At zero field, we observe two characteristic resonances corresponding to the $T_y$-$T_z$ and $T_x$-$T_z$ transitions. Generally, for a spin-1 system we expect a Hamiltonian of the form

$$H = \hbar D \left( S_z^2 - \frac{2}{3} \right) + \hbar E \left( S_x^2 - S_y^2 \right) - \hbar \gamma_{el} \vec{S} \cdot \vec{B} \quad (1),$$

where $\hbar$ denotes the reduced Planck constant, $D$ and $E$ the zero-field splitting parameters (ZFS), $\gamma_{el} = (2\pi) \times 28$ GHz/T the gyromagnetic ratio of an electron, $\vec{S}$ the vector of Pauli spin matrices for a spin-1 system, and $\vec{B}$ the magnetic field. Since the experiments are performed on a randomly oriented ensemble of EYFP molecules, we expect the ODMR signal to be an average over all molecular orientations, resulting in a powder spectrum [36]. Fitting at low magnetic fields (the region where the uncertainty in the field calibration is small, Methods), we extract the ZFS parameters $D = (2\pi) \times (2.356 \pm 0.004)$ GHz and $E = (2\pi) \times (0.458 \pm 0.003)$ GHz. *Ab initio* calculations with the B3LYP-optimized geometry and $\omega$B97X-D3 functional predict similar values, $D_{DFT} = (2\pi) \times 2.42$ GHz and $E_{DFT} = (2\pi) \times 0.57$ GHz (SI Table 5). Using the fitted D and E parameters, we calculate the expected spectra of the powder-averaged transition frequencies as a function of magnetic field (Fig. 2b). The overall magnetic-field dependent broadening and asymmetric resonance profile is well captured by our model. It also extrapolates well to higher fields exemplified at 34.1 mT (Fig. 2c bottom).

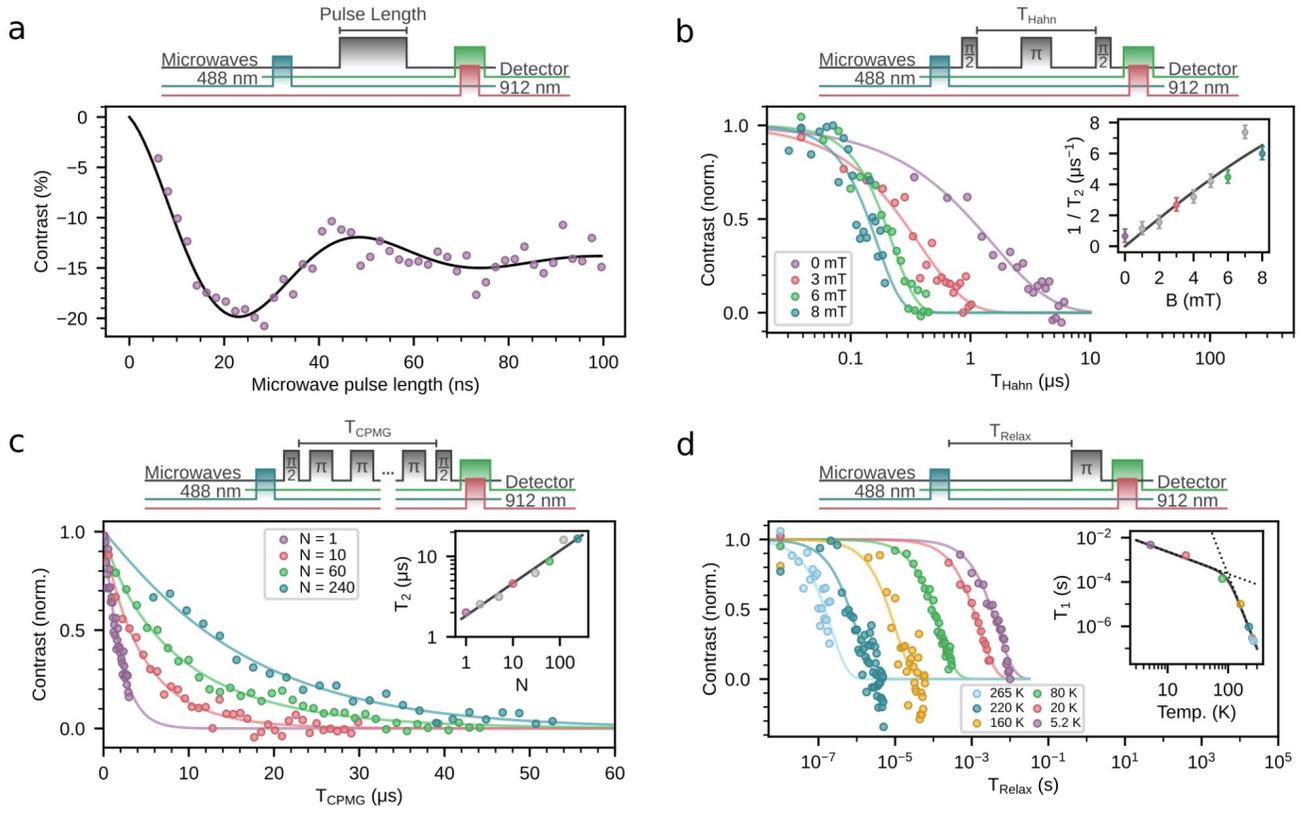

**Fig. 3 | Coherent control of EYFP qubits. a,** Rabi oscillations of EYFP $T_x$-$T_z$ transition driven at a frequency of 2.815 GHz. The fit corresponds to an exponentially damped cosine (black). **b,** Spin coherence as a function of total evolution time ($T_{Hahn}$) under Hahn-echo decoupling for different magnetic fields, with stretched-exponential fits. Inset shows the scaling of the Hahn-echo dephasing rate ($1/T_2$) as a function of external magnetic field and the corresponding theoretical model (black). **c,** Same as in **b**, but under a CPMG decoupling sequence for different numbers of $\pi$-pulses (N) and no external field. The inset shows $T_2$ scaling as a function of N and its corresponding fit (black). **d,** Contrast as a function of evolution time ($T_{Relax}$) for different sample temperatures fit to exponential decays. Inset shows the spin-lattice relaxation time ($T_1$) as a function of temperature fitted with equation 2, with fitted amplitudes A = (43±8) $K^{-1}s^{-1}$ and B = (47±7)×$10^{-12}$ $K^{-7}s^{-1}$. In all plots, gray data points indicate additional measurements shown in the inset but not the main plot for clarity. The fit errors in the insets of panels **c** and **d** are smaller than the data points and omitted. Panels **a-c** are measured at 80 K.

**Qubit coherence**

Driving the EYFP molecules with a fixed frequency that is resonant with the $T_x$-$T_z$ transition results in coherent Rabi oscillations between the spin sublevels (Fig. 3a). The oscillations are highly damped, with a decay rate that increases with Rabi frequency as a result of random molecular orientations and microwave drive detuning (Extended Data Fig. 6).

We measure the spin coherence time ($T_2$) using a Hahn-echo sequence to refocus dephasing caused by quasi-static magnetic fields (Fig. 3b). With no applied field, the $T_2$ is (1.5±0.2) μs. We find that with increasing field, $T_2$ decreases to (140±30) ns at 7 mT. The scaling of $T_2$ is a consequence of the $T_x$-$T_z$ resonance resembling a clock-like transition [37, 38], resulting in a dephasing rate $1/T_2 \propto \gamma_{eff} \sim \frac{\gamma B_z}{\sqrt{(E/\gamma_{el})^2 + B_z^2}}$. This rate is in good agreement with our observations (Fig. 3b inset), suggesting that, in the vicinity of the clock transition, qubit dephasing remains primarily limited by magnetic rather than electric field noise.

Applying a CPMG sequence improves $T_2$ by shifting the dynamical decoupling filter function to higher frequencies [39]. With 240 $\pi$-pulses, we observe a $T_2^{DD}$ of (16±2) μs at zero field (Fig. 3c), which represents a 15-fold enhancement compared to the $T_2$ determined using Hahn Echo. Repeating the CPMG measurement for different numbers of $\pi$-pulses (N) yields a scaling $T_2 \sim N^{0.40 \pm 0.03}$ (Fig. 3c inset), which provides insights into the magnetic noise properties and can be connected to a power spectral density $S(\omega) \propto \omega^{-2/3}$ [40].

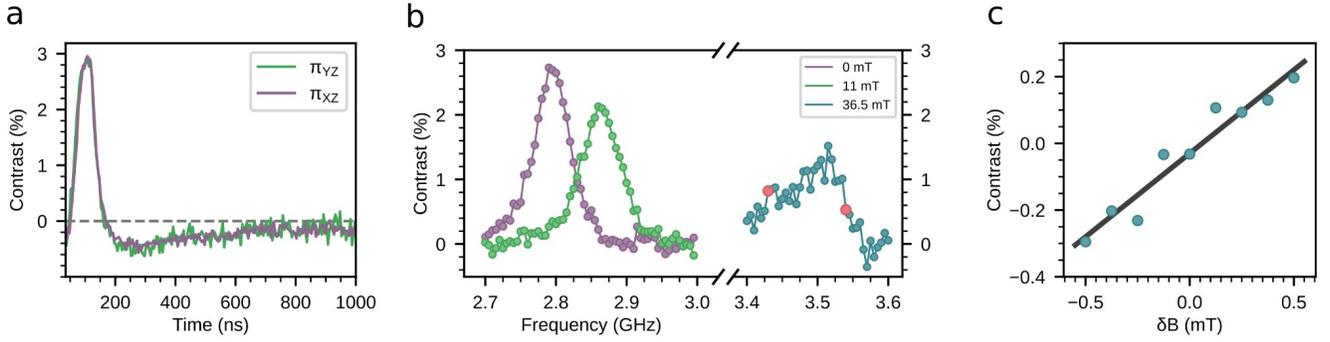

**Fig. 4 | Room temperature quantum sensing. a,** Contrast of OADF-based spin readout as a function of time after the 912 nm laser pulse for the $T_y$-$T_z$ (green) and $T_x$-$T_z$ (purple) transition. **b,** ODMR in aqueous solution at various magnetic fields. Red points indicate the locations of the field sensing experiment in **c**. **c,** DC magnetic field sensing operating at a bias field of 36.5 mT by measuring the difference in the ODMR signal at the two frequencies indicated in **b**. Each data point has been averaged for 15 min. Panels **a-c** are measured at room temperature.

We characterize the spin-lattice relaxation time and its dependence on temperature by measuring the spin contrast between the qubit initialized in $T_z$ and $T_x$ as a function of wait time. At 80 K, the spin polarization follows an exponential decay with a lifetime of (141±5) μs (Fig. 3d). For an integer spin system, the depolarization rate is given by $1/T_1 = AT + BT^7$ (2), where $A$ and $B$ are the amplitudes of direct and Raman processes, respectively [36] (Fig. 3d inset). This model provides a good fit to our data when the amplitudes are treated as free parameters, but unambiguous identification of the exact relaxation mechanisms will require a more thorough investigation.

**Operation at room temperature**

Despite the fast spin relaxation extrapolated from our fit, we observe ODMR of EYFP in aqueous solution at room temperature with a contrast of nearly 3% (Fig. 4a, 4b). Interestingly, the sign of the spin contrast is reversed compared to measurements performed at 80 K, which suggests that spin-dependence of the ISC and/or RISC rates are temperature dependent. We confirm the magnetic origin of the resonance by repeating the experiment in the presence of an external field (Fig. 4b).

We demonstrate the potential of our qubit as a room-temperature DC magnetic field sensor following an approach adapted from the atomic clock community [41]. Detecting subtle shifts in the ODMR spectrum by measuring at two frequency points, one above and one below the $T_x$-$T_z$ resonance (Fig. 4b, right), results in a contrast difference that is a linear function of a small external magnetic field ($\delta B$) (Fig. 4c). From this measurement, we can estimate a DC magnetic field sensitivity of 98 pT mol$^{½}$ Hz$^{-½}$ (Methods). The presence of a clock transition at zero field requires us to operate at a bias magnetic field.

**Qubit sensor expressed in mammalian cells**

We highlight the potential of *in vitro* applications of our genetically-encodable qubit by performing spin measurements on EYFP expressed in human embryonic kidney (HEK) 293T cells (Methods). Fluorescence imaging of the cells adhered to the sapphire substrate confirms that the protein remains localized within the cells (Fig. 5a). After cooling to 175 K, confocal fluorescence scans of the cells correlate well with the room-temperature widefield images (Extended Data Fig. 9). Measuring over six bright regions containing cells, we record an ODMR signal (Fig. 5b, red). The OADF associated with these scans correlates well with the cells observed in widefield and confocal fluorescence images. Measuring only the pixels outside of cells results in a dim background signal, indicating that the spin resonance signature originates from the bright cells rather than any extracellular EYFP. Repeating the same process, we drive Rabi oscillations in the cells (Fig. 5c).

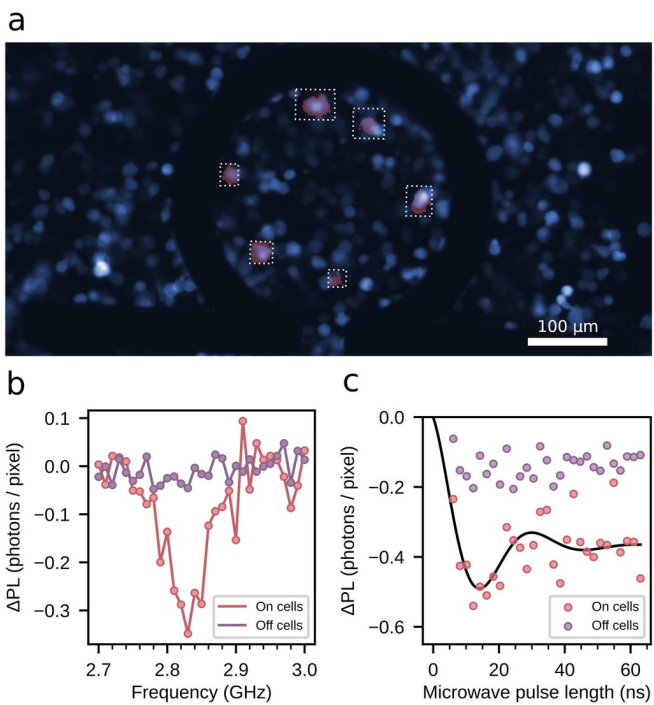

**Fig. 5 | Coherent control in cells. a,** A widefield fluorescence image of HEK 293T cells expressing cytosolic EYFP (blue). A loop structure for applying the microwave drive is visible. ODMR scans are taken in the white outlined regions, and thresholded for the brightest pixels (red). **b,** ODMR signal averaged over the bright pixels (red, on cells) and dark pixels (purple, off cells). Instead of contrast, the PL difference normalized to the number of pixels is shown in order to illustrate the amplitude of the signal contributed by the background. **c,** Rabi signal taken with the same process as in **a**, **b**. Panel **a** widefield image is measured at room temperature. Red pixel overlay in **a**, and panels **b**, **c** are measured at 175K.

The ODMR and Rabi signals are consistent with those of the purified protein, confirming that the spin and optical interface are robust to operation within a cell.

**Discussion and outlook**

Our results establish fluorescent proteins as a genetically-encodable spin qubit platform with readout contrast, coherence, and spin-lattice relaxation times comparable to solid-state color centers. We estimate that EYFP has an upper-bounded AC sensitivity of 183 fT mol$^{1/2}$ Hz$^{-1/2}$ at liquid-nitrogen temperatures (Methods) and an upper-bounded DC sensitivity of 93 pT mol$^{1/2}$ Hz$^{-1/2}$ at room temperature. For reference, a proton spin located 5 nm away from the qubit – a practical distance for an EYFP fusion protein [42] – would generate a magnetic field of approximately 20 nT at the qubit's location. For a proton polarization $p$, these parameters translate into a proton number sensitivity of $\frac{1}{p}80$ pmol Hz$^{-1}$ (Methods). While this sensitivity falls short of state-of-the-art NV diamond sensors [43], the ability to utilize the large existing library of genetically-encoded fusion proteins for *in vitro* and *in vivo* experiments provides significant advantages over diamond sensors.

Enhancements in sensitivity can be achieved by improving coherence or spin readout. At liquid-nitrogen temperatures, coherence is limited by magnetic-field noise, which can be drastically reduced by substituting nearby proton spins with deuterium, resulting in expected coherence times approaching 100 μs [44, 45]. Spin readout can be enhanced through two complementary strategies. First, initialization of the qubit requires an efficient population transfer from the singlet ground to the metastable triplet state. This can be improved by optimizing the excitation conditions and reducing unwanted RISC caused by the initialization laser [33, 46]. Second, OADF readout yields at most one photon per molecule. Combining OADF readout with fluorescence cycling in the singlet or triplet manifold could increase the photon counts by 300-fold ($1/\phi_{T1}$, where $\phi_{T1}$ denotes the triplet yield [33, 34]). Optimizing the imaging optics could further improve per-molecule collection efficiency by a factor of 810 (Methods). Collectively, these improvements would yield at least a 500-fold improvement in AC and DC sensitivity. Finally, in addition to the above approaches for increasing the readout signal, combining our system with existing single-molecule microscopy techniques could enable the detection of individual EYFP qubit sensors [47], paving the way to single-molecule spin experiments.

Our on-demand readout method unlocks a large class of metastable triplet systems as qubits. In addition to EYFP, optically-activated RISC has been observed in many fluorescent proteins and dye molecules [34, 35]. In fact, systems comprising a singlet ground and excited state coupled to a metastable triplet are ubiquitous among fluorescent molecules, many of which may possess the requisite photophysical parameters for their metastable triplet to be addressed using our spin-readout technique. The exploration of these molecules may be accelerated by the computational methods demonstrated here for predicting the optical transitions and ZFS parameters.

The quantum information science and bioengineering fields have developed a diverse array of approaches to engineering complex systems. For example, the creation of highly regular structures through molecular self-assembly [48] could allow for the engineering of one-, two-, and three-dimensional paramagnetic molecular arrays [49]. The many-body dynamics of these spin systems could be optically read out with a protein qubit. Furthermore, traditional quantum engineering approaches to improving optical properties and spin coherence rely on gaining an understanding of qubit physics from first principles. Genetically-encodable probes like EYFP can be engineered using directed evolution, a black-box optimization approach that uses high-throughput screening of protein variants. Directed evolution of fluorescent proteins has resulted in a vast array of different spectral properties and functions, which highlight their tunability and rich photophysics [50]. Directed evolution on our EYFP qubit could be used to optimize its optical and spin properties, and even reveal unexpected insights into qubit physics. Protein-based qubits are positioned to take advantage of techniques from both quantum information sciences and bioengineering, and produce transformative advances in both fields.

**Acknowledgments** We thank Dr. Martin Di Federico and Dr. Gustavo Cancelo for their efforts in the development of the QICK platform for spin qubits, Prof. Melody Swartz and Prof. Jotham Austin for help with cell imaging, Dr. Maria del Mar Ingaramo and Dr. Andrew York for insightful discussions related to the photophysics of fluorescent proteins and for sharing alternative proteins showing a magnetic-field effect [51], Prof. Stefan Stoll for discussion related to time-resolved EPR spectroscopy, and Dr. Oleg Poluektov and Dr. Jens Niklas for measuring time-resolved EPR spectroscopy on similar proteins. We thank Dr. Jonathan Marcks, Dr. Junghyun Lee, Prof. Christopher Anderson, Cyrus Zeledon, Dr. William Grubbe, and Shannon Lu for illuminating discussions, and Xiaofei Yu, Uri Zvi, Sanskriti Chitransh, and Lingjie Chen for initial measurements on metastable triplet states. J.S.F., B.S.S., S.V., Z.Z.G., E.G.R., A.E.K., D.D.A., and P.C.M. acknowledge financial support from the NSF QuBBE QLCI (NSF OMA- 2121044). In addition, J.S.F., D.D.A., and P.C.M. acknowledge support through the Moore Foundation grant number 12216. Z.Z.G. received funding through NIH 75N93024C00023-0-9999-1 & 75N93019C00041-P00014-9999-1. We acknowledge the use of the Pritzker Nanofabrication Facility at the University of Chicago (NSF ECCS-2025633) and the The University of Chicago Biophysics Core Facility (RRID:SCR_017915).

**Author contributions**
J.S.F. and B.S.S. constructed the experimental setup and performed the measurements. S.V. carried out the DFT calculations. Z.Z.G. expressed the EYFP in mammalian cells. J.S.F., B.S.S., S.W., B.K., and E.G.R. performed simulations. J.S.F., B.S.S., Y.T., L.R.W., D.D.A., and P.C.M. developed the experimental design. J.S.F., B.S.S., and M.X. developed the bacterial protein expression and purification process. J.S.F., B.S.S., S.W., E.G.R., Y.T., L.R.W., D.D.A., and P.C.M. analyzed the data. J.S.F., B.S.S., D.D.A., and P.C.M. wrote the manuscript with contributions from the coauthors. J.H., A.E.K., L.G., D.D.A., and P.C.M. supervised the work.

**Competing interests**
J.S.F., B.S.S., D.D.A., and P.C.M. are inventors on patent application no. 63/712,179 submitted by the University of Chicago that covers fluorophore-based spin qubits and associated methods.